\documentstyle[prd,aps]{revtex}
\preprint{
\text{LAVAL-PHY-97-14},
\text{hep-ph}
}
\begin{document}
\author{L. Marleau}
\address{D\'epartement de Physique, Universit\'e Laval\\
Qu\'ebec, Canada, G1K 7P4}
\title{Deformed Skyrmions}
\maketitle

\begin{abstract}
The spherically symmetric hedgehog ansatz used in the description of the
skyrmion is believed to be inadequate for the rotational states such as the
nucleon ($I=J=\frac{1}{2}$) and the $\Delta $ ($I=J=\frac{3}{2}$) due to
centrifugal forces. We study here a simple alternative: an oblate spheroidal
solution which leads to lower masses for these baryons. As one might expect,
the shape of the solution is flatter as one increases $I=J$ whether the size
of the soliton is allowed to change or not.
\end{abstract}

\section{Introduction}

When Skyrme first introduced its model a few decades ago \cite{Skyrme61} to
describe baryons as solitons in a non-linear field theory of mesons, the
solution proposed was a spherically symmetric hedgehog ansatz. There are
reasons to believe that this solution is not adequate for the rotational
states such as the nucleon ($I=J=\frac{1}{2}$) and the $\Delta $ ($I=J=\frac{%
3}{2}$) due to centrifugal forces \cite{centrifugal}. Several treatments
have been proposed in the past with relative success \cite
{deformed,deformed2}. In this work, we take a naive approach and propose a
simple alternative. Instead of the spherically symmetric hedgehog solution,
we introduce an oblate spheroidal solution. This leads to lower masses and
quadrupole deformations for these baryons. Moreover, the shape of the
solution is flatter as one increases $I=J$ whether one allows the size of
the soliton to change or not.

\section{The static oblate soliton}

The oblate spheroidal coordinates $\left( \eta ,\theta ,\phi \right) $ are
related to Cartesian coordinates through 
\begin{equation}
(x,y,z)=d(\cosh \eta \sin \theta \cos \phi ,\cosh \eta \sin \theta \sin \phi
,\sinh \eta \cos \theta ).  \nonumber
\end{equation}
A surface of constant $\eta $ corresponds to a sphere of radius $d$
flattened in the $z$-direction by a factor of $\tanh \eta $. For $\eta $
small, the shape of the surface is more like that of a pancake of radius $d$
whereas for large $\eta $, one recovers a spherical shell of radius $r=\frac{%
de^{\eta }}{2}$.

We would like to replace the hedgehog solution for the Skyrme model by an
oblate solution. Writing the Lagrangian for the Skyrme model \cite{Adkins83}
(neglecting the pion mass) 
\begin{equation}
{\cal L}=-\frac{F_{\pi }^{2}}{16}Tr\left( L_{\mu }L^{\mu }\right) +\frac{1}{%
32e^{2}}Tr\left( \left[ L_{\mu },L_{\nu }\right] ^{2}\right)
\end{equation}
where $L_{\mu }=U^{\dagger }\partial _{\mu }U$ with $U\in SU(2)$. Let us now
define a static oblate solution by 
\begin{equation}
U=e^{i\left( {\bf \tau \cdot \hat{\eta}}\right) f\left( \eta \right) }\ 
\label{oblate}
\end{equation}
where$\ {\bf \hat{\eta}}$ is the unit vector$\ {\bf \hat{\eta}}=\frac{{\bf %
\nabla }\eta }{\left| {\bf \nabla }\eta \right| }$. The boundary conditions
for the winding number $N=1$ solution are $f\left( 0\right) =\pi $ and $%
f\left( \infty \right) =0$. Note that this is not a priori a solution of the
field equations derived from the Skyrme Lagrangian.

Integrating over the angular variables $\theta $ and $\phi ,$ the static
energy reads 
\[
E_{s}=\epsilon \int d\eta \left[ \frac{\widetilde{d}}{2}\left( \alpha
_{21}f^{\prime 2}+\alpha _{22}\sin ^{2}f\right) +\frac{1}{4\widetilde{d}}%
\left( \alpha _{41}f^{\prime 2}\sin ^{2}f+\alpha _{42}\sin ^{4}f\right)
\right] 
\]
with 
\[
\begin{tabular}{ll}
$\alpha _{21}\left( \eta \right) =2\cosh \eta $ & $\alpha _{22}\left( \eta
\right) =2\left( -2\cosh \eta +\left( 2\cosh ^{2}\eta -1\right) L\left( \eta
\right) \right) $ \\ 
$\alpha _{41}\left( \eta \right) =2L\left( \eta \right) $ & $\alpha
_{42}\left( \eta \right) =\frac{1}{2}\left( \frac{1}{\cosh ^{2}\eta }\left(
2\cosh \eta +L\left( \eta \right) \right) +\frac{2\cosh \eta }{\left( \cosh
^{2}\eta -1\right) }\right) $%
\end{tabular}
\]
where $L\left( \eta \right) \equiv \ln \left( \frac{\cosh \eta +1}{\cosh
\eta -1}\right) $ and the constants are defined by $\epsilon =\frac{2\sqrt{2}%
\pi F_{\pi }}{e}$ and $\widetilde{d}\equiv \frac{eF_{\pi }}{2\sqrt{2}}d.$

Minimizing the static energy with respect to $f\left( \eta \right) $, we
then solve numerically the corresponding non-linear ordinary second-order
differential equation. For calculational purposes, we set the value of the
parameters of the Skyrme model as $F_{\pi }=129$ MeV, $e=5.45$ (and $m_{\pi
}=0$) which coincide with those of ref. \cite{Adkins83} obtained by fitting
for the masses of the nucleon and the $\Delta $ in the hedgehog ansatz.

The solution near $\eta \rightarrow 0$ has the form $f\left( \eta \right)
\sim \pi -a_{1}\eta ,$ whereas in the limit $\eta \rightarrow \infty $ one
recovers the spherical symmetry with $f\left( \eta \right) \sim \frac{k}{%
(de^{\eta })^{2}}$ where $a_{1}$ and $k$ are constants which depend on $%
\widetilde{d}$.

The masses of the nucleon and of the $\Delta $-isobar get contributions both
from the static and rotational energy and will generally depend on the
choice of $\widetilde{d}$. We fix the value of $\widetilde{d}$ for each
baryon by minimizing its mass with respect to $\widetilde{d}$.

\section{Collective variables}

Using the oblate solution, we then compute the masses of the nucleon and of
the $\Delta $-isobar. However, when one departs from the spherical symmetry
of the hedgehog ansatz, it is customary to introduce extra collective
variables for isorotation in addition to those characterizing spatial
rotation since these are no longer equivalent, in general. The spin and
isospin contributions to the rotational energy are however equal in our case
since we use solution (\ref{oblate}) and we are only interested in ground
states with ${\bf K}={\bf J}+{\bf I}=0$. As a result, we need only consider
one set of collective variables.

We work in the body-fixed system and assume that the time dependence can be
introduced using the usual substitution $U\rightarrow A(t)UA^{\dagger }(t)$
where $A(t)$ is a time-dependent $SU(2)$ matrix. We can then go on and treat 
$A(t)$ approximately as quantum mechanical variables.

The quantization procedure is fairly standard and leads to principal moments
of inertia $I_{11}$ and $I_{33}$ in the body-fixed system. We get a
representation analog to a symmetrical top with the rotational kinetic
energy in space and isospace 
\begin{equation}
E_{rot}^{J,J_{3}}=\frac{1}{2I_{11}}\left( \frac{\left| {\bf J}\right|
^{2}+\left| {\bf I}\right| ^{2}}{2}\right) +\frac{1}{2}\left( \frac{1}{I_{33}%
}-\frac{1}{I_{11}}\right) J_{3}^{2}.  \label{Bodyfixed}
\end{equation}
where $\left| {\bf J}\right| ^{2}$ and $\left| {\bf I}\right| ^{2}$ are the
spin and the isospin respectively and, $J_{3},$ the $z$-component of the
spin. We have already used the relation $J_{3}=-I_{3}$ here which follows
from axial symmetry of the ansatz. Added to the static energy $E_{s}, $ it
leads to the total energy $M^{J,J_{3}}=E_{s}+E_{rot}^{J,J_{3}}$ identified
with the mass of the baryon.

Observables states, however, must be eigenstates of $\left| {\bf J}\right|
^{2},$ $J_{3},$ $\left| {\bf I}\right| ^{2}$, $I_{3}$ with eigenvalues $%
J(J+1),$ $m_{J}$, $I(I+1),$ $m_{I}$ where the operators now refer to the
laboratory system (as opposed to body-fixed operators in (\ref{deformed})
and above). These eigenstates are taken into account by direct products of
rotation matrices 
\begin{equation}
\left\langle {\bf \Omega }|J,m_{J},m\right\rangle \left\langle {\bf \omega }%
|I,m_{I},-m\right\rangle =D_{m_{J}m}^{J}\left( {\bf \Omega }\right)
D_{m_{I}-m}^{I}\left( {\bf \omega }\right)  \label{WignerD}
\end{equation}
where ${\bf \Omega }$ and ${\bf \omega }$ are, respectively, the Euler
angles for the rotation and isorotation from the body-fixed frame to the
laboratory system. The explicit calculation of the energy of rotation
requires in general the diagonalization $E_{rot}^{J,J_{3}}$. (see ref. \cite
{deformed} for more details)

The minimization of the static energy for the spherical symmetric ansatz
gives $E_{s}=8.20675\epsilon $, $M_{N}=8.906\epsilon $ and $M_{\Delta
}=11.703\epsilon $. For the oblate spheroidal ansatz, the parameter $%
\widetilde{d}$ is chosen in order to minimize the mass of the corresponding
baryon. In general, as $\widetilde{d}$ increases, the static energy $E_{s}$
deviates from its lowest energy value given by the spherical hedgehog
configuration. On the other hand, oblate configurations have larger moments
of inertia which tend to decrease the rotational kinetic energy. The
existence of a non-trivial oblate spheroidal ground state for the nucleon
and the $\Delta $-isobar, as it turns out, depends mostly on the relative
importance of static and rotational energy.

We find that the ground state for the nucleon is almost spherical but
nonetheless oblate with $\widetilde{d}=0.0013$ thus exhibiting a small
quadrupole deformation and a slightly lower mass with respect to a spherical
configuration. For the $\Delta $-isobar, the oblateness or quadrupole
deformation is even more important and accounts for a $4\%$ decrease in
mass. We obtain a minimum mass for a value of $\widetilde{d}=0.32$ with $%
M_{\Delta }=11.293\epsilon $.

Since the minimum of the ground state is affected by the oblate shape of the
solution, the parameters $F_{\pi }$ and $e$ as given in ref. \cite{Adkins83}
no longer reproduce the quantities they were designed to fit. We must
readjust $F_{\pi }$ and $e$ which determine the value of $\widetilde{d}$ for
the nucleon and $\Delta $-isobar respectively. After several iterations, we
find $F_{\pi }=118.4$ MeV and $e=5.10$ with $\widetilde{d}=0.0014$ ($%
\widetilde{d}=0.40$) for the nucleon ($\Delta $-isobar).

\section{Discussion}

Quadrupole deformations were found previously in the context of rotationally
improved skyrmions. Contrary to our approach, these solutions involve the
minimization of an Hamiltonian which also includes the (iso) rotational
kinetic energy. Yet, we found that the oblate spheroidal ansatz gives lower
energy than the spherical one for baryon ground states. Of course, ansatz (%
\ref{oblate}) is not necessarily the lowest energy solution.

It may also be interesting to consider deformations of the oblate skyrmions
under scaling of the unitary transformations $U\left( {\bf r}\right) $ such
that $U\left( {\bf r}\right) =U_{0}\left( \rho {\bf r}\right) $ to minimize
the total energy of the nucleon and $\Delta $-isobar. The total energies $%
M_{N}(\rho )$ and $M_{\Delta }(\rho )$ can be minimized with respect to the
scaling parameter $\rho $, i.e. to the energically favored size of the
oblate skyrmion. The energies are found to be $M_{N}(\rho _{\min
}=0.868)=8.797\epsilon $ and $M_{\Delta }(\rho _{\min
}=0.670)=10.064\epsilon $ for both the oblate case compared with $M_{N}(\rho
_{\min }=0.867)=8.799\epsilon $ and $M_{\Delta }(\rho _{\min
}=0.668)=10.238\epsilon $ for spherical ansatz. The baryon ground states are
now swelled oblate solutions. Again, one should in principle readjust the $%
F_{\pi }$ and $e$ parameters to fit the masses of the nucleon and $\Delta $%
-isobar.

This work was done in collaboration with F. Leblond. Support from the NSERC
of Canada and by the FCAR du Qu\'{e}bec is acknowledged.

\end{document}